\documentstyle[twocolumn,aps,psfig,floats]{revtex}  

\begin{document} 

\def\be{\begin{equation}}
\def\ee{\end{equation}}

\draft  

\twocolumn[\hsize\textwidth\columnwidth\hsize\csname %
@twocolumnfalse\endcsname

\title{Phase-Field Formulation for Quantitative Modeling of Alloy Solidification} 

\author{Alain Karma}

\address{
Physics Department and Center for Interdisciplinary Research
on Complex Systems, \\
Northeastern University, Boston, Massachusetts 02115 
}

\date{March 7, 2001}

\maketitle

\begin{abstract}        
A phase-field formulation is introduced 
to simulate quantitatively microstructural pattern formation in alloys.
The thin-interface limit of this formulation yields a much less
stringent restriction on the choice of interface thickness than previous
formulations and permits to eliminate non-equilibrium effects at the interface.
Dendrite growth simulations with vanishing solid
diffusivity show that both the interface
evolution and the solute profile in the  
solid are well resolved.  
\end{abstract} 

]

The phase-field approach has
emerged as a method of choice to simulate 
microstructural evolution during
solidification 
\cite{pfgen,pfalloy,KarRap,Alm,McFetal,Proetal,PlaKar,Kimetal2}. 
This approach has the well-known advantage
that it avoids to explicitly track a sharp boundary
by smearing the interface region over some thickness $\sim W$.
However, it has the disadvantage that it is
hard to use quantitatively. 
This is because it is often computationally  
too stringent to choose $W$ small enough
to resolve the desired sharp-interface limit
of the phase-field model, even on computers of today. 
This is especially true for small growth rates where the 
scale of the microstructure is typically several orders of magnitude
larger than the microscopic capillary length $d_0$. 

Progress has recently been made
to overcome this difficulty by using a ``thin-interface''
analysis of the phase-field model \cite{KarRap,Alm,McFetal} 
where $W$ is assumed small compared to 
the scale of the pattern,
but not smaller than $d_0$. 
For the standard symmetric model (with equal 
thermal conductivities in the solid and liquid),
Karma and Rappel (KR) have shown 
that this thin-interface limit yields   
two essential benefits \cite{KarRap}. 
Firstly, it maps onto the standard set of 
sharp-interface equations that one obtains
in the classical sharp-interface limit where 
$W/d_0\rightarrow 0$, but yields a much less stringent
restriction on $W/d_0$ that renders the computations tractable. 
Secondly, it makes it possible to eliminate interface kinetic effects
by a specific choice of phase-field model parameters. 

These two
properties combined have made this thin-interface limit
ideally suited to model dendritic growth in pure materials
quantitatively at low undercooling when 
used in conjunction with efficient
numerical algorithms \cite{Proetal,PlaKar}. However,
how to extend these results in a useful way
to the more general case where the two phases 
do not have symmetrical properties has 
remained an open challenge.  
Using a distinct thin-interface 
analysis, Almgren showed that the 
two-sided model with unequal thermal conductivities 
maps onto a modified set of sharp-interface equations that is
plagued by finite interface thickness effects \cite{Alm}. 
These include (i) a 
temperature jump across the interface, (ii) an
interface stretching correction to the 
heat conservation condition at the interface
(Stefan condition) also found previously in \cite{FifPen},
and (iii) a surface diffusion correction to the
same condition.  

The same effects plague the thin-interface limit of 
existing phase-field models
of alloy solidification \cite{pfalloy,Kimetal2} 
and make them inadequate to 
model quantitatively the low growth rate
regime of experimental relevance. 
In alloys, (i) translates into
a chemical potential jump across the interface
associated with the well-known effect 
of solute trapping \cite{KurFis},
and (ii) and (iii) modify 
the mass conservation condition at the interface. 
Here, I present a phase-field formulation
of alloy solidification that makes it possible to eliminate
\emph{simultaneously} all three effects. Furthermore, I demonstrate
that it yields the same computational benefits as the thin-interface
limit of the symmetric model \cite{KarRap}  
for dendritic growth.  

For clarity of exposition, I first discuss the thin-interface
limit of a simpler model 
that describes an idealized binary alloy with 
parallel liquidus and solidus slopes, and which
reduces to the standard Hele-Shaw flow 
problem \cite{SafTay} in its Laplace limit. 
I then consider a realistic dilute alloy model with
unequal slopes that reduces
to the former model in the limit where the partition
coefficient $k\rightarrow 1$.  
The equations of the first model are
\begin{eqnarray}
& &\tau\partial_t \phi  =
W^2\nabla^2\phi\,
- \partial_\phi \left[ f(\phi)+\lambda\,g(\phi)\,u \right],
\label{p1}\\
& &\partial_t c\,+\,\vec \nabla \cdot \vec j=0,
 \label{p2}
\end{eqnarray}
where $c$ is the concentration defined as the
mole fraction of B in a binary alloy of A and B,
\begin{equation}
\vec j \,=\,-\Delta c_0\, D\,q(\phi)\vec\nabla u\,
-\,a_t\,W\,\Delta c_0\,\partial_t\phi \,
\vec \nabla \phi/|\vec \nabla\phi|,
\label{cur}
\end{equation}
and 
\begin{equation}
u\equiv c/\Delta c_0+h(\phi)/2-(c_{s0}+c_{l0})/(2\Delta c_0)
\end{equation}
is a dimensionless measure of the departure 
of the chemical potential from its
equilibrium value with $u=0$ in equilibrium.    
$c_{s0}$ ($c_{l0}$) are the equilibrium concentrations in
the solid (liquid) at a fixed temperature $T_0$,
$\Delta c_0\equiv c_{l0}-c_{s0}$,  
$f(\phi)=-\phi^2/2+\phi^4/4$ is a double well
function with the minima $\phi=\pm 1$ corresponding to the solid ($+1$)
and liquid ($-1$), $g(\phi)$ is an odd function of $\phi$
with $g(\pm 1)=\pm 1$ and vanishing first and second
derivatives $g'(\pm 1)=g''(\pm 1)=0$, and $h(\phi)$ is an odd
function of $\phi$ with $h(\pm 1)=\pm 1$ that can 
be chosen independently of $g(\phi)$ for
non-variational dynamics \cite{KarRap}.  

The above model reduces 
to the symmetric model for the choice $q(\phi)=1$ 
and $a_t=0$. In this case, the  
thin-interface limit of this model maps onto the  
Stefan problem defined by:
$\partial_tu=D\nabla^2u$ in both phases, the Stefan
condition $V=-D(\partial_n u|^+-\partial_n u|^-)$, 
where $V$ is the interface
velocity and $\partial_n u|^\pm$ is the  
normal gradient of $u$ on the solid $(-)$ and liquid
side $(+)$ of the interface, and the velocity-dependent
Gibbs-Thomson condition
\begin{equation}
u=-d_0\kappa-\beta V, \label{gt}
\end{equation}
where 
\begin{equation}
d_0=a_1W/\lambda,
~~{\rm and}~~ \beta=a_1\left[\tau/(W\lambda)-a_2W/D\right] \label{kin}.
\end{equation}
The expressions for 
the coefficients $a_1$ and $a_2$ are identical to those
derived by KR \cite{KarRap} and interface kinetics
can be eliminated ($\beta=0$) by choosing
$\lambda=D\tau/(a_2W^2)$.

Next, for the alloy case, 
$q(\phi)$ must now be chosen 
to vary from $q(-1)=1$ in the liquid to
$q(+1)=D_{\rm solid}/D$ in the solid. I consider explicitly
the one-sided limit where $D_{\rm solid}/D\rightarrow 0$,
but the results also extend to the more realistic
case where $D_{\rm solid}/D\ll 1$.
The essential new term that
yields the desired thin interface limit is the 
\emph{anti-trapping} mass current that
corresponds to the second term on the 
right-hand-side of equation (\ref{cur}), and
is only non-vanishing in the diffuse interface region.
It produces a solute flux from the solid to the liquid along
the direction normal to the interface that
counter-balances the trapping current associated with
the jump of chemical potential across the interface:
$\Delta u\equiv u^+-u^-$, where $u^\pm$ 
correspond to the liquid ($+$)
and solid ($-$) sides of the interface.
Thus the anti-trapping current makes it possible to eliminate
this jump while still leaving enough freedom  
to choose the other functions in the model to eliminate the corrections
to the mass conservation condition.    
Repeating the analyses of KR \cite{KarRap} and
Almgren \cite{Alm},
I obtain that $\Delta u$ vanishes if 
$F^+=F^-$, where
\begin{equation}
F^\pm\,=\, \int_0^{\pm\infty} d\eta \left[p(\phi_0(\eta))-p(\mp 1)\right],
\end{equation}
and
\begin{equation}
p(\phi_0)=\left(h(\phi_0)-1
+a_t\sqrt{2}\,(1-\phi_0^2)\right)/q(\phi_0).
\end{equation}
In the above definitions, $\phi_0(\eta)=-\tanh(\eta/\sqrt{2})$ is
the equilibrium phase-field profile,
where $\eta$ is a coordinate that runs normal to the
interface scaled by $W$, and I have used the identity
$\partial_\eta\phi_0=-(1-\phi_0^2)/\sqrt{2}$. Next, the
mass conservation condition has the form
\begin{equation}
V=-D\,\partial_n u|^+ - \, c_1 W \kappa V  - \, c_2 W D \,\partial_s^2 u
\label{modcons}
\end{equation}
where $c_1\equiv H^+-H^-$, $c_2\equiv Q^+-Q^-$, and
\begin{eqnarray}
H^\pm &=&\int_0^{\pm\infty} d\eta \left[h(\phi_0(\eta))-h(\mp 1)\right],\\
Q^\pm &=&\int_0^{\pm\infty} d\eta \left[q(\phi_0(\eta))-q(\mp 1)\right].
\end{eqnarray}
The second and third term on the
right-hand-side of equation (\ref{modcons}) 
represent the solute redistribution  
due to stretching the interface and by
diffusion along its arclength $s$, respectively.
These two terms appear, equivalently, 
at two successive orders in the
thin interface limit considered by
KR, or both at second order in
the distinct thin-interface limit
considered by Almgren \cite{Alm}. 

In summary, one is left with three conditions to satisfy: 
(i) $F^+=F^-$ (chemical potential jump), 
(ii) $H^+=H^-$ (stretching), and (iii) 
$Q^+=Q^-$ (surface diffusion). It is actually 
possible to satisfy two of these three conditions simultaneously
within the standard phase-field formulation without
the anti-trapping current.
For example, with $a_t=0$ in equation (\ref{cur}), 
the choice $h(\phi)=1-(1-\phi)^2/2$  
and $q(\phi)=(1-\phi)/2$ satisfy (i) and (iii),
but not (ii), and it is also possible to satisfy
(i) and (ii) but not (iii). However, test simulations 
show that with either $c_1$ or $c_2$ non-vanishing,  
the morphological stability of the interface that sets
the initial scale of the pattern 
becomes significantly altered for computationally
tractable choices of $W$ \cite{Folchetal}, which is also easy to
check analytically by repeating the standard Mullins-Sekerka
analysis with the modified boundary condition (\ref{modcons}).
Similarly, it can be shown that 
solvability theory predictions for the dendrite
tip become modified. Consequently, all three 
conditions must be satisfied to lift 
the restriction on $W$.

With the anti-trapping current present, one is
now free to make the simplest choices
$h(\phi)=\phi$ and $q(\phi)=(1-\phi)/2$ that satisfy $H^+=H^-$ and
$Q^+=Q^-$, respectively. By choosing $a_t=1/(2\sqrt{2})$, 
one can then reduce the function $p(\phi_0)$ to 
the simple form $p(\phi_0)=\phi_0-1$ that
satisfies $F^+=F^-$ (and a non-vanishing amount of trapping can
also be obtained by varying $a$).
Remarkably, with these choices, the thin interface limit  
of the one-sided model produces a velocity-dependent
Gibbs-Thomson condition that is 
\emph{identical} to the
one of the symmetric model. Consequently the 
expressions for $a_1$ and $a_2$ that determine
$d_0$ and $\beta$ are the same as in Ref. \cite{KarRap}:
$a_1=I/J$ and $a_2=(K+JF)/(2I)$ where
$I=\int_{-\infty}^{+\infty}d\eta (\partial_\eta\phi_0)^2$,
$J=g(-1)-g(+1)$, $F\equiv F^\pm =\sqrt{2}\ln 2$, and  
\begin{equation}
K=\int_{-\infty}^{+\infty}d\eta \,\partial_\eta\phi_0(\eta)
g'(\phi_0(\eta)) 
\int_0^\eta d\xi\, \phi_0(\xi), \label{Kform}
\end{equation}
where I have defined $g'(\phi)\equiv \partial_\phi g(\phi)$. 
It also follows that the standard  
Hele-Shaw flow problem with an infinite viscosity contrast \cite{SafTay}
can be simulated by taking the limit $\partial_tu \rightarrow 0$
of the present phase-field model. 

Consider now the standard 
one-sided dilute alloy model defined by
the set of equations
\begin{eqnarray}
\partial_t c&=&D\nabla^2c,\label{fb1} \\
c_l(1-k)V&=&-D\partial_nc|^+,\label{fb2}\\
c_l/c_l^0&=&1-(1-k)d_0\,\kappa \label{fb3}
\end{eqnarray} 
where $d_0=\gamma T_M/[L|m|(1-k)c_l^0]$ is the
chemical capillary length, $T_M$ is the melting
temperature, $L$ is the latent heat of melting, 
$m$ is the liquidus slope, $k=c_s/c_l$
is the partition coefficient where $c_l$ ($c_s$) is
the concentration on the liquid (solid) side of
the interface, and 
solidification is again assumed to 
take place isothermally. 

To construct a  
thin interface limit that maps onto the free-boundary
problem (\ref{fb1})-(\ref{fb3}), I follow the
same procedure of adding a local 
anti-trapping current in order to 
eliminate the jump of chemical potential together
with the other terms.
The equations are  
\begin{equation}
 \tau\partial_t \phi  =
W^2\nabla^2\phi\,
- \partial_\phi \left[ f(\phi)+\frac{\lambda}{1-k}\,g(\phi)(e^u-1) \right],
\label{pa1} 
\end{equation}
with the same continuity relation 
(\ref{p2}) as before,
\begin{eqnarray}
\vec j &=&-D\,c\,q(\phi)\vec\nabla u\,-
\,a_t\,W\,c_l^0 (1-k) \,e^u\,\partial_t\phi \,
\vec \nabla \phi/|\vec \nabla\phi|,
\label{cur2}\\
u&=&\ln\left[\frac{2c/c_l^0}{1+k-(1-k)h(\phi)}\right],\label{u2}
\end{eqnarray}
and where $g(\phi)$, $h(\phi)$, and $q(\phi)$ obey
the same limits at $\phi=\pm 1$.
$u$ is again a dimensionless measure of
the departure of the chemical potential $\mu$
from equilibrium  
(namely here $u\equiv v_0(\mu-\mu_E)/(RT_0)$ where $R$ is the
rare gas constant and $v_0$ is the molar volume
assumed to be constant). 
The logarithmic dependence
of $u$ on $c$
is related to the entropy
of mixing in the free-energy density as in
previous models \cite{pfalloy,Kimetal2}. 
The main difference here is again the addition of
the anti-trapping mass current. 
Also, the present 
thin-interface analysis differs from the analysis of
Kim {\it et al.} \cite{Kimetal2} that does not consider  
interface stretching and surface diffusion, and
assumes that $q(\phi)$ is constant 
in the interface region.
 
The condition for eliminating surface diffusion 
becomes now $Z^+=Z^-$, where 
$Z^\pm =\int_0^{\pm\infty} d\eta \left[M(\eta)-M(\pm\infty)\right]$
where $M(\eta)\equiv q(\phi_0(\eta))\,c_0(\eta)$, and
\begin{equation}
c_0(\eta)=c_l^0\left[1+k-(1-k)h(\phi_0(\eta))\right]/2
\end{equation} 
is the equilibrium concentration profile across 
a stationary interface. Therefore choosing
\begin{equation}
q(\phi)=\frac{1-\phi}{1+k-(1-k)h(\phi)} \label{qphi}
\end{equation} 
satifies
this condition. The choices
$h(\phi)=\phi$ and $a_t=1/(2\sqrt{2})$ then make
the jump of $u$ and the interface
stretching term vanish 
up to finite corrections that
turn out to be small in computations and
will be discussed elsewhere. Consequently,  
the conditions for $c$ on the two
sides of the interface have the desired form,
$c_l/c_l^0=1-(1-k)d_0\kappa-(1-k)\beta V$ and
$c_s=kc_l$ where the expressions for $d_0$, $\beta$, 
$a_1$, and $a_2$ are again identical to
those for the symmetric model quoted earlier here.
Therefore, $\beta$ can again be made to vanish.
The dilute alloy model is easily shown to reduce to the parallel slope
model in the limit $k\rightarrow 1$ 
by making the change of variable $U=(e^u-1)/(1-k)$.  
A small solid diffusivity can also be modeled by adding
$(1+\phi)D_{\rm solid}/(2D)$ to the right-hand-side of 
equation (\ref{qphi}), which has a negligible
effect on the thin-interface limit in typical alloys where
solid diffusion is substitutional 
($D_{\rm solid}/D \sim 10^{-4}$) \cite{KurFis}.

Finally, the model is straightforward to extend to directional
solidification with the standard frozen temperature
approximation $T(z)=T_0+G(z-V_pt)$ that yields
(for $\beta=0$) the interface condition 
\begin{equation}
c_l/c_l^0=1-(1-k)d_0\kappa-(1-k)(z-V_pt)/l_T,
\end{equation}
where $V_p$ is the pulling velocity of the
sample along the $z$-axis, $l_T=|m|(1-k) c_l^0/G$
is the thermal length , $G$ is the temperature
gradient, and $c_l^0=c_\infty/k$
where $c_\infty\equiv c(z=+\infty)$.
It simply suffices to substitute $e^u$ by
$e^u+(1-k)(z-V_pt)/l_T$ in equation (\ref{pa1}).

The convergence of the model was examined by carrying out   
two-dimensional simulations of isothermal dendritic growth.
These simulations are directly analogous to 
the ones carried out previously to test the 
thin interface limit of the symmetric model \cite{KarRap}.
Crystalline anisotropy was included
by generalizing equation (\ref{pa1}) to a standard
anisotropic form 
\begin{eqnarray}
& &\tau(\Theta)\,\partial_t \phi=
- \partial_\phi \left[ f(\phi)+\frac{\lambda}{1-k}\,g(\phi)(e^u-1) \right]
 \label{pp1} \\
& &+\,{\vec \nabla}\cdot(W(\Theta)^2{\vec \nabla} \phi)
-\partial_x\left(W(\Theta)W'(\Theta)\partial_y\phi\right)\nonumber\\
& &
+\,\partial_y\left(W(\Theta)W'(\Theta)\partial_x\phi\right),
\nonumber 
\end{eqnarray}
where $\Theta\equiv \arctan(\partial_y \phi/\partial_x\phi)$
is the angle between the direction normal to the phase-field 
interface and the $x$-axis. As a result, 
the anisotropic form of $d_0$ and $\beta$ become  
$d_0(\theta)=a_1[W(\theta)+W''(\theta)]/\lambda$
and $\beta(\theta)=a_1\left[\tau(\theta)/(W(\theta)\lambda)
-a_2W(\theta)/D\right]$ in the interface condition where
$a_1=0.8839$, and $a_2=0.6267$ for the common
choice $\partial_\phi g(\phi)=(1-\phi^2)^2$.
Furthermore, I chose the standard form of
four-fold anisotropy $W(\theta)=W a_s(\theta)$
with $a_s(\theta)=1+\epsilon_4 \cos 4\theta$ and made $\beta(\theta)$
vanish by letting $\tau(\theta)=\tau a_s(\theta)^2$ and
$\lambda=D\tau/(a_2W^2)$.  

I simulated equations (\ref{p2}) and
(\ref{pp1}) with $\vec j$ and $u$ defined 
by equations (\ref{cur2}) and (\ref{u2}), respectively.
I compare the results of the present model  
with $a_t=1/(2\sqrt{2})$ and $q(\phi)$ defined by equation (\ref{qphi}),
to the more standard choice (i.e. similar to
previous models) that has no anti-trapping current
($a_t=0$) and uses the simplest scaled diffusivity function
$q(\phi)=(1-\phi)/2$; $h(\phi)=\phi$ and $\partial_\phi g(\phi)=(1-\phi^2)^2$
in both models. The former model has
the desired thin-interface limit with 
local equilibrium at the interface, whereas the latter has
both a chemical potential jump and surface diffusion.
I used a simple finite-difference
Euler method with $\Delta x=0.4$ and
$\Delta t=0.008$,
$W=\tau=1$, $\epsilon_4=0.02$, $k=0.15$, and
the scaled supersaturation 
$\Omega=(c_l^0-c_\infty)/[c_l^0(1-k)]=0.55$ where $c_\infty$
is the initial alloy concentration. In all simulations,
the initial condition
consisted of a circular seed of radius $r=22\, d_0$,
$u=\ln (1-(1-k)\Omega)$, and $c$ defined by equation (\ref{u2})
that varies from $c_\infty$ in liquid to
$kc_{\infty}$ in solid. 

\begin{figure}
\centerline{
\psfig{file=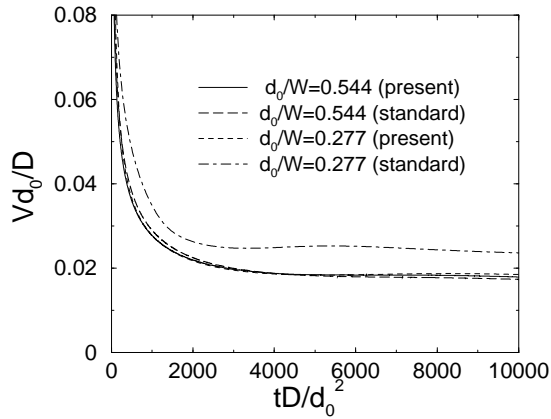,width=.4\textwidth}}
\smallskip
\caption{Plots of scaled dendrite tip
velocity $Vd_0/D$ vs scaled time $tD/d_0^2$ for 
$a_t=1/(2\sqrt{2})$ and $q(\phi)$ given
by Eq. \ref{qphi} (present), and $a_t=0$
and $q(\phi)=(1-\phi)/2$ (standard).}
\label{fig1}
\end{figure}

\begin{figure}
\centerline{
\psfig{file=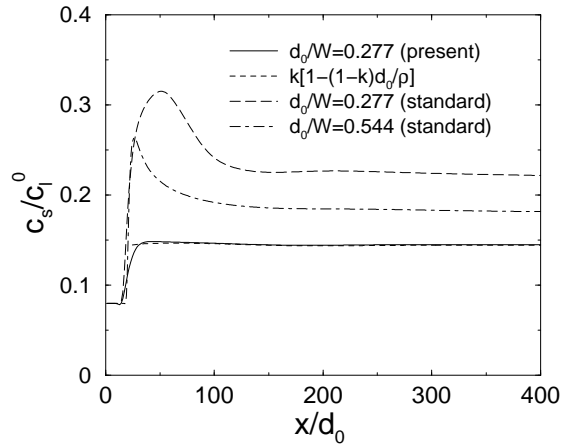,width=.4\textwidth}}
\smallskip
\caption{Plots of solute profiles in the solid along the 
central dendrite axis, and comparison with the
Gibbs-Thomson relation for the present model
(short-dash line).}
\label{fig2}
\end{figure}

The dimensionless dendrite
tip velocity $Vd_0/D$ is plotted vs the dimensionless 
time $tD/d_0^2$ for the two models and two different
ratios of $d_0/W$ in Fig. \ref{fig1}. Since
the simulation time scales $\sim (d_0/W)^5$, the runs
with $d_0/W$ twice smaller are $\approx 32$ times shorter.
Furthermore, I have plotted in 
Fig. \ref{fig2} the scaled concentration $c_s(x)/c_l^0$
in the solid vs the scaled position $x/d_0$ 
along the central dendrite 
axis for the two different models. 
Plots in Figs. \ref{fig1}-\ref{fig2} 
must superimpose when results are converged.
These plots show, as expected,
that the present model is well converged in
this range of $d_0/W$ that is comparable
to the one studied in the 
symmetric model \cite{KarRap}, whereas the standard
model is not. This is especially true for the 
microsegregation profile that is still far from being converged 
in the latter model, even
for the largest ratio $d_0/W=0.544$. In contrast, 
this profile is already well converged
for a twice smaller ratio in the present model; it agrees,
self-consistently, within a few percent
with the Gibbs-Thomson relation
$c_s(x)/c_l^0=k[1-(1-k)d_0/\rho]$ where $\rho$
is the dendrite tip radius in the simulation.
It will be shown elsewhere that 
this dramatic difference of convergence for 
microsegregation is due to
the fact that the amount of solute trapped
$\sim p \ln p$ for small velocity
($p=WV/D\ll 1$) in the standard model.

The present results demonstrate that the
phase-field method can be successfully extended
to model quantitatively microstructural pattern 
formation in alloys with a realistic solid diffusivity.
For this important application, it is potentially
more advantageous than the level set 
method \cite{levelset} since it does not require
the explicit computation of the interface velocity. 
These results also revive
the hope to extend the phase-field method to model 
accurately a wide range of
other interfacial patterns
with a strong asymmetry between phases.

This research is supported by U.S. DOE Grant 
No. DE-FG02-92ER45471 and benefited from 
computer time allocation at NERSC and NU-ASCC.
I thank Roger Folch and Mathis Plapp for
valuable discussions.

\end{document}